\documentclass[a4paper,12pt]{article}
\usepackage{newlfont}
\usepackage{amscd,amsfonts,amsmath,amssymb,amstext,amsthm,latexsym}
\begin{document}
\newcommand{\Ad}{\mathrm{Ad}}
\newcommand{\ad}{\mathrm{ad}}
\newcommand{\tr}{\mathrm{tr}}
\newcommand{\re}{\mathrm{Re}}
\newcommand{\acts}{\mbox{ \raisebox{0.27ex}{\tiny{$\bullet$}} }}
\newcommand{\sacts}{\,\mbox{\raisebox{-0.5ex}{\Large{$\cdot$}}}}
\newcommand{\ssacts}{\,\mbox{\raisebox{-0.4ex}{\Large{$\cdot$}}}}
\newcommand{\sodass}{\ \vert \ }
\newcommand{\inv}{^{-1}}
\newcommand{\ddt}{\frac{\mathrm d}{\mathrm d t}\biggl\vert_{t=0}}
\newcommand{\seq}[1]{({#1})_{n \in \nn}}
\newcommand{\sperp}{{\mbox{\tiny {$\perp$}}}}
\newcommand{\mgc}{{\mg}^{\mbox{\tiny $\cc$}}}
\newcommand{\munc}{\mun^{\mbox{\tiny $\cc$}}}
\newcommand{\kc}{K^\mathbb C}
\newcommand{\gx}{{G \acts x}}
\newcommand{\gv}{{G \acts v}}
\newcommand{\kbeta}{{K \acts \beta}}
\newcommand{\obeta}{\mathcal O_\beta}
\newcommand{\ip}{{i\mathfrak p}}
\newcommand{\ia}{i \mathfrak a_+^\ast}
\newcommand{\iao}{i \mathfrak a_0^\ast}
\newcommand{\tma}{\tilde {\mathfrak a}}
\newcommand{\mip}{{\mu_{i \mathfrak p}}}
\newcommand{\mia}{\mu_{i \mathfrak a}}
\newcommand{\tmia}{\mu_{i \tilde {\mathfrak a}}}
\newcommand{\mipa}{{\mu_{\ip,\alpha}}}
\newcommand{\mipb}{{\mu_{\ip,\beta}}}
\newcommand{\dmip}{\mathrm d \mip(x)}
\newcommand{\mipn}{\mu_\ip^{(n)}}
\newcommand{\Mip}{\mathcal M_{i \mp}}
\newcommand{\Mipb}{\mathcal M_{\ip,\beta}}
\newcommand{\tMip}{\tilde {\mathcal M}_\ip}
\newcommand{\SGMip}[1]{\mathcal S_{G}(\Mip({#1}))}
\newcommand{\SGMipb}[1]{\mathcal S_{G}(\Mipb({#1}))}
\newcommand{\tSGMip}{\mathcal S_{G}(\tilde{\mathcal M}_\ip)}
\newcommand{\SGM}{\mathcal S_{G}(\Mip)}
\newcommand{\SAM}{\mathcal S_A(\mathcal M_{i \ma})}
\newcommand{\zcone}{\mathcal N_{G}}
\newcommand{\zconess}{\mathcal N_{G_s}}
\newcommand{\Vohne}{V \backslash \{0\}}
\newcommand{\xa}{X_{\alpha}}
\newcommand{\xb}{X_{\beta}}
\newcommand{\xc}{X_{(T_\gamma)}}
\newcommand{\Xa}{X^{T_\alpha}}
\newcommand{\pn}{{\mathbb P^n(\mathbb C)}}
\newcommand{\hyp}[1]{\mathcal O_{#1}}
\newcommand{\zs}{Z_\beta^{\mbox{\raisebox{0.3ex}{\scriptsize $ss$}}}}
\newcommand{\zss}{Z_\beta^{\mbox{\raisebox{0.3ex}{\scriptsize $ss$}}} {\mbox{\raisebox{0.2ex} {\scriptsize$(+)$}}}}
\newcommand{\zplus}{Z_\beta{\mbox{\raisebox{0.2ex} {\scriptsize$(+)$}}}}
\newcommand{\ma}{\mathfrak a}
\newcommand{\mb}{\mathfrak b}
\newcommand{\mg}{\mathfrak g}
\newcommand{\mh}{\mathfrak h}
\newcommand{\mk}{\mathfrak k}
\newcommand{\ml}{\mathfrak l}
\newcommand{\mm}{\mathfrak m}
\newcommand{\mn}{\mathfrak n}
\newcommand{\mo}{\mathfrak o}
\renewcommand{\mp}{\mathfrak p}
\newcommand{\mq}{\mathfrak q}
\newcommand{\mr}{\mathfrak r}
\newcommand{\ms}{\mathfrak s}
\newcommand{\mt}{\mathfrak t}
\newcommand{\mun}{\mathfrak u}
\newcommand{\mz}{\mathfrak z}
\newcommand{\cc}{\mathbb C}
\newcommand{\rr}{\mathbb R}
\newcommand{\zz}{\mathbb Z}
\newcommand{\nn}{\mathbb N}
\newcommand{\pp}{\mathbb P}
\newcommand{\qd}{\hfill{\( \square \)}}
\newcommand{\bn}{\bigskip \noindent}
\newcommand{\m}{\medskip \noindent}
\newcommand{\n}{\noindent}
\newfont{\mib}{cmmib10}
\newtheorem{theorem}{Theorem}[section]
\newtheorem{proposition}{Proposition}[section]
\newtheorem{lemma}{Lemma}[section]
\newtheorem{corollary}{Corollary}[section]
\newtheorem{definition}{Definition}[section]
\newtheorem{remark}{Remark}[section]
\newtheorem{example}{Example}[section]
\newtheorem{conjecture}{Conjecture}[section]
\newtheorem{fact}{Fact}

\title{Level dynamics and the ten-fold way}

\author{A. Huckleberry$^a$, M. Ku\'s$^b$, P. Sch\"utzdeller$^a$
\\
\\
$^a$Fakult\"at f\"ur Mathematik, Ruhr-Universit\"at Bochum,\\
D-44780 Bochum, Germany
\\
\\
$^b$Centrum Fizyki Teoretycznej PAN, \\
Al. Lotnik\'ow 32/42, 02-668 Warszawa,
Poland}

\maketitle

\abstract We investigate the parameter dynamics of eigenvalues of Hamiltonians
('level dynamics') defined on symmetric spaces relevant for condensed matter
and particle physics. In particular we: 1) identify appropriate reduced
manifold on which the motion takes place, 2) identify the correct Poisson
structure ensuring the Hamiltonian character of the reduced dynamics, 3)
determine the canonical measure on the reduced space, 4) calculate the
resulting eigenvalue density.

\section{Introduction}
The concept of statistical approach to parametric level dynamics proved to be
very fruitful in explaining the applicability of the Random Matrix Theory to
the statistics of spectra of generic quantum systems \cite{haake00}. In the most
straightforward setting it consists in considering the flow in a (sub)space $Q$ of
$N\times N$ Hermitian matrices
\begin{equation}\label{flow}
X\mapsto X+tY,
\end{equation}
where $Y$ is a constant, Hermitian, $N\times N$ matrix and $t$ is a real
parameter. The matrix $X_t=X+tY$ represents here the Hamiltonian of the quantum
system in question, where $X$ and $Y$ describe the `unperturbed' and
'perturbing' parts, respectively, and $t$ - a coupling parameter controlling
the strength of the perturbation. Depending on the symmetries of the
investigated system \cite{haake00}, $X$ and $Y$ are general, complex Hermitian
matrices, real symmetric matrices, or Hermitian matrices fulfilling
$X=JX^tJ^{-1}$ where
\begin {gather}\label {complex structure}
J=
\begin {pmatrix}
0 & -I\\
I & 0
\end {pmatrix}
.
\end {gather}

To gain information about various statistical properties of the distribution of
eigenvalues of $X_t$ we should be able to deduct from (\ref{flow}) the
parametric motion of them. This is achieved by treating $Q$ as a configuration
space of a Hamiltonian motion in the phase space $Q\times Q$ with the
fictitious time $t$ and reducing the dynamics to a smaller manifold which on which
the motion is still Hamiltonian and the eigenvalues of $X_t$ are explicitly
used as coordinates. The resulting equations can be interpreted as describing
the dynamics of a fictitious gas of interacting particles to which one applies
rules of statistical mechanics, in. particular, in search of the `equilibrium'
distribution of the particle positions (ie., in fact, the eigenvalues of
$X_t$).

The general reduction procedure was explained in \cite{hzkh01}. Recently it
became obvious that besides the above enumerated symmetry classes of
Hamiltonians there are other ones, relevant for condensed matter and particle
physics \cite{zirnbauer96,heinzner05}, where $Q$ is a symmetric space. It is
thus of considerable interest to extend the investigations of the parametric
level dynamics to these cases. To achieve the goal we should 1) identify
appropriate reduced manifold on which the motion takes place, 2) identify the
correct Poisson structure ensuring the Hamiltonian character of the reduced
dynamics, 3) determine the canonical measure on the reduced space, 4)
calculate the resulting eigenvalue density. The above enumerated partial goals
will be completed in the consecutive section of the paper.

Let us start with a general description of the setting, and let $G/K$ be
one of the following symmetric spaces of non-compact type

\m
$SU(m,n)/S(U(m)\times U(n)),$

\m
$SO(m,n)^0/(SO(m) \times SO(n)),$

\m
$Sp(2m,2n)/Sp(m) \times Sp(n),$

\m
$SL(n,\rr)/SO(n),$

\m
$SL(n,\mathbb H)/Sp(n),$

\m
$SO^\ast(2n)/U(n),$

\m
$Sp(n,\rr)/U(n)$.

\bn
The configuration space of the considered dynamics of the type (\ref{flow}) is
then identified with one of the above.

Note that in every of the above cases there exists a closed embedding of $G_0$
into $SL_N(\cc)$ for some $N \in \nn$ such that the image is a closed subgroup
of $SL_n(\cc)$ which is closed under conjugate transpose inverse, given as the
common zero set of some set of real-valued polynomials in the real and
imaginary parts of the matrix entries. In the following we will only consider
this image in $SL_N(\cc)$ which we also denote by $G_0$. Then $K$ is the
fixed point set of the Cartan involution, which is given by $g \mapsto
(g^\dagger)\inv$ and therefore a subgroup of $SU(N)$.

Let $\mg_0$ and $\mk_0$ be the Lie algebras of $G_0$ and $K$, respectively.
The Cartan decomposition of $\mg_0$ is given by $\mg_0 = \mk_0 \oplus \mp_0$
where $\mk_0$ is the $+1$-eigenspace of the Cartan involution $X \mapsto
-X^\dagger$ and $\mp_0$ the corresponding $-1$ eigenspace. Then the symmetric
space $G_0/K$ can be identified with $\mp_0$. The corresponding phase space
is then the cotangent bundle of $G_0/K$ and can be identified with $\mp_0
\times \mp_0^\ast$. Using the Killing form on $\mg_0$ which is given by $(X,Y)
\mapsto \re (\tr (XY))$ we can further identify the cotangent bundle with $\mp
\times \mp$. The standard symplectic structure of the cotangent bundle then
induces the symplectic structure $ \omega = \re(\tr(\mathrm d X \land \mathrm d
Y) $ on $\mp \times \mp$.

Let now $\ma$ be a maximal Abelian subspace of $\mp$ and define $S:= \ma \times
\mp = \ma \times \ma \times \ma^\sperp$, where $\ma^\sperp$ is the orthogonal
complement of $\ma$ in $\mp$ with respect to the inner product $B\vert_{\mp
\times \mp}$. Let $(q,p,r)$ denote the linear coordinates on $S$ corresponding
to the decomposition $S = \ma \times \ma \times \ma^\sperp$. Since $\mp = K
\acts \ma$ the action map $\alpha : K \times S \to \mp \times \mp, (k,\xi)
\mapsto k \acts \xi$ is surjective and $S$ is a slice for the $K$ action on
$N$. The stabilizer of $K$ in a generic point $x_0 \in S$ is the centralizer
$M$ of $\ma$ in $K$, $M = Z_{K}(\ma)$. Since $M$ need not to be trivial,
$S$ is in general not an exact slice for the $K$-action. To define an exact
slice we need an exact slice for the $M$-action on $\ma^\sperp$.

\m For the last four symmetric spaces the group $M$ is trivial, so $S = \ma
\times \ma \times \ma^\sperp$ is an exact slice, whereas for the first three
ones the group $M$ is non trivial and even non-abelian. In the following we
compute explicitly the exact slice for the case of the symmetric space
$SU(p,q)/S(U(p) \times U(q))$ and then we show how this allows to determine the
exact slices for the other two nontrivial cases of $SO(m,n)^0/S(O(m) \times
O(n))$ and $Sp(2m,2n)/Sp(m) \times Sp(n)$.

\section{Computing the exact slice}

\m
The Cartan decomposition of $\mg_1 = \ms\mun(m,n), m \geq n$, is given by
$k_1 \oplus \mp_1$ with
 \[
\mk_1 = \left \{
 \begin{pmatrix}
 A & 0 \cr
0 & D
\end{pmatrix}
\vert \  A \in \mun(m), D \in \mun(n)
\mbox{ and } tr(A) + tr(D) = 0 \right \}
= \ms( \mun(m) \times \mun(n)),
\]
and
\[
\mp_1 = \left \{
\begin{pmatrix}
0 & B \cr
\bar B^T & 0
\end{pmatrix} \vert  \ B \in M_{m \times n}(\cc)
\right \}.
\]
As a maximal abelian subspace of $\mp_1$ we take
\[
\ma_1 = \left  \{ X=
\begin{pmatrix}
0 & B \cr
\bar B^T & 0
\end{pmatrix}
\mbox{ with } B=
\begin{pmatrix}
\vdots & \cdots & \vdots \cr
0 & \cdots & 0 \cr
0 &  \cdots & a_n \cr
0 & \cdot & 0  \cr
a_1 & \cdots & 0
\end{pmatrix},
\ a_j \in \rr \right\}.
 \]
Then we can compute the centralizer of $\ma_1$ in
$K_1 = S (U(m) \times U(n))$. It is given by
\[
M_1:= Z_{K_1}(\ma_1) = \biggl \{
\begin{pmatrix}
U & 0\cr
0 & T
\end{pmatrix} \in K_1
\ \vert \  U  \in U(m-n)
 \]
and
 \[
\qquad  \qquad \qquad  T= diag
(e^{it_1},\dots,e^{it_n},e^{it_n},\dots,e^{it_1}) \biggr \}.
\]
The restricted roots with respect to $\ma_1$ are given by
$\pm2 f_i$, $\pm f_i$ and  $\pm f_i \pm f_j$ for $i \neq j$,
where $f_i \in \ma_1^\ast$ is given by
 \[
f_i : \ma_1 \to \rr, \quad X \mapsto a_i.
 \]
We choose a notion of positivity on this restricted roots such that
$f_i, 2f_i, f_i +f_j$ and $f_i -f_j$  with $i <j$ are positive.
The corresponding restricted  root spaces are given in \cite{knapp05}
p. 371 Example 2. The dimensions of the restricted root spaces read
\begin{center}
$\dim \mg_{2f_i} = 1$, \quad  $ \dim \mg_{f_i} = m-n$ \quad  and \quad
$ \dim \mg_{f_i \pm f_j } = 2$.
\end{center}
Consider the map
\[
(1- \theta) : \mg_1 \to \mp_1, \quad X \mapsto X- \theta(X),
\]
where $\theta$ is the Cartan involution for $\mg_1^\cc = \ms\ml(m+n,\cc)$, so
$\theta(X) = - \bar X^T$. In particular every restricted root space
$\mg_\alpha$ provides a subspace $(1-\theta)(\mg_\alpha)$ of $\mp_1$. The group
$M_1$ acts on each of this spaces and we compute the slice for the action of
$M_1$ on $\ma_1^\sperp$ by analyzing the action of $M_1$ on
$(1-\theta)(\mg_\alpha)$.


We start with the restricted roots $f_i$. The space $(1-\theta)(\mg_{f_i})$ is
of the form
\[
\left \{
\begin{pmatrix}
0 &  0 &  R \cr
0 & 0 & 0 \cr
\bar R^T & 0 &  0 \cr
\end{pmatrix}
\Biggr \vert \ R =
\begin{pmatrix}
0 & \cdots & 0 & v_1 & 0 & \cdots & 0 \cr
0 & \cdots & 0 & \vdots & 0 & \cdots & 0 \cr
0 & \cdots & 0 & v_{m-n} & 0 & \cdots & 0 \cr
\end{pmatrix}
\in M_{m-n,n}(\cc) \right\} \simeq \cc^{m-n}.
\]
\[
\begin{matrix}
\uparrow \cr
i
\end{matrix}
\]
The action of $M_1$ on $(1-\theta)(\mg_{f_i})$ is given by the
standard representation of
\[
\left \{ \begin{pmatrix} U & 0 \cr 0 & 0 \end{pmatrix} \in M_1 \right\}
\simeq U(m-n)
\]
 on $\cc^{m-n}$. We will see later that this subgroup acts trivially on the
images of the other root spaces, so we can compute the slice separately. The
slice for the standard representation of $U(m-n)$ on
\[
\bigoplus_{i = 1}^n (1- \theta)(\mg_{f_i}) \simeq \underbrace{\cc^{m-n} \times
\cdots \times \cc^{m-n}}_{n-times}
\]
is given by
\[
\left \{
\left( \left(
\begin{array}{c}
c_1 \\
0 \\
\vdots \\
0 \\
\end{array}\right) , \left(
\begin{array}{c}
\ast \\
c_2 \\
 0 \\
\vdots  \\
\end{array}\right) , \dots , \left(
\begin{array}{c}
\ast \\
\vdots  \\
c_{n-1} \\
0 \\
\end{array}\right) , \left( \begin{array}{c}
 \ast\\
 \vdots\\
 \ast\\
c_n
\end{array}\right), \left( \begin{array}{c}
 \ast\\
 \vdots\\
 \ast\\
\ast
\end{array}\right) , \dots , \left( \begin{array}{c}
 \ast\\
 \vdots\\
 \ast\\
\ast
\end{array}\right) \right) \Biggl \vert \  c_j  \in \rr^+ \right\}.
\]
The group $M_1$ acts trivially on the spaces $(1- \theta)(\mg_{2 f_i})$ which
are of the form
\[
\left \{
\begin{pmatrix}
0 &  0 & 0 \cr
0 & 0 & T \cr
0 & \bar T^T &  0 \cr
\end{pmatrix}
\Biggr \vert \ T = antidiag( 0,\dots 0, ia_i, 0 , \dots , 0)
\in M_{n,n}(\cc) \right\}.
\]
The spaces $(1-\theta)(\mg_{f_i \pm f_j})$ are of the form
\[
\left \{
\begin{pmatrix}
0 &  0 & 0 \cr
0 & 0 & S \cr
0 & \bar S^T &  0 \cr
\end{pmatrix}
\Biggr \vert \ S \in M_{n,n}(\cc)  \mbox{ with } s_{ij} =  \pm \ \bar s_{ji}
\in \cc\right\} \simeq \cc.
\]
The group $M_1$ acts on $(1-\theta)(\mg_{f_i \pm f_j}) \simeq \cc$ by $ z
\mapsto e^{it_i -it_j}z$, so the slice for this action is $\rr^+$. Actually we
can get this slice for the root spaces corresponding to simple roots which are
given by $f_i \pm f_{i+1}$. Hence the slice for $M_1$ on $\ma_1^\sperp$
consists of all matrices of the form
\[
\begin{pmatrix}
\begin{matrix}
0 & \cdots & 0 &  \ \cr
\vdots &  \ddots &  \vdots  & \ \cr
0 & \cdots & 0 &  \ \cr
\cr
\cr
\end{matrix}
&
\begin{matrix}
\quad 0 & 0 & \hdots & \hdots & 0 & 0 & \quad  \vline \cr
\quad \vdots  & &  &  &  & \vdots & \quad \vline \cr
\quad 0 & 0 & \hdots & \hdots  & 0 & 0 & \quad \vline \cr
& & & & & & \quad \vline \cr
 \cr
\end{matrix}
&
\begin{matrix}
\diamond & \ast &  \ast & \ast  &  \cdots &\ast \cr
& \ddots & \ast  & \ast  & \cdots &\ast\cr
0 &  &\diamond &  \ast & \cdots & \ast\cr
\cr
\hline \cr
\end{matrix}
\cr
\begin{matrix}
0 & \cdots & 0 &  \ \cr
\vdots &  \ddots &  \vdots  & \ \cr
0 & \cdots & 0 &  \ \cr
\vdots &  \ddots &  \vdots  & \ \cr
0 & \cdots & 0 &  \ \cr
\cr
\hline
\end{matrix}
&
\begin{matrix}
\quad 0 & 0 & \hdots & \hdots & 0 & 0 & \quad \vline \cr
\quad \vdots  & &  &  &  & \vdots &\quad \vline \cr
\quad 0 & 0 & \hdots &\hdots & 0 & 0 & \quad \vline \cr
\quad \vdots  & &  &  &  & \vdots & \quad \vline \cr
\quad 0 & 0 & \hdots & \hdots  & 0 & 0 & \quad \vline \cr
& & & & & & \quad \vline \cr
\hline
\end{matrix}
&
\begin{matrix}
\ast & \ast & \ast & \ast &\diamond & i a_1 \cr
\ast & \ast & \ast & \diamond & \cdot & \diamond \cr
\ast & \ast & \diamond & \cdot  &\diamond  & \ast \cr
\ast & \diamond &  \cdot &\diamond  &\ast  & \ast \cr
\diamond & \cdot & \diamond & \ast  &\ast  & \ast \cr
i a_n & \diamond & \ast & \ast & \ast &\ast  \cr
\cr
\hline
\end{matrix}
\cr
\begin{matrix}
 & & & \vline\cr
\diamond & 0 & 0 & \vline \cr
  \ast & \ddots & 0 & \vline \cr
 \ast & \ast & \diamond & \vline\cr
 \ast & \ast & \ast & \vline\cr
 \vdots & \vdots & \vdots & \vline \cr
 \ast & \ast & \ast & \vline
\end{matrix}
&
\begin{matrix}
& & & & & &  \vline\cr
 \ast & \ast & \ast & \ast &\diamond & -i a_1& \vline \cr
\ast & \ast & \ast & \diamond & \cdot & \diamond  & \vline\cr
\ast & \ast & \diamond & \cdot  &\diamond  & \ast & \vline \cr
\ast & \diamond &  \cdot &\diamond  &\ast  & \ast & \vline\cr
\diamond & \cdot & \diamond & \ast  &\ast  & \ast & \vline \cr
-i a_n & \diamond & \ast & \ast & \ast &\ast  & \vline\cr
\cr
\end{matrix}
&
\begin{matrix}
\cr
\quad 0 & 0 & 0 & \cdots  & 0 & 0 &  \cr
 \quad \vdots  & &  &  &  & \vdots & \cr
\quad 0 & 0 & 0 & \cdots  & 0 & 0 & \cr
\quad 0 & 0 & 0 & \cdots & 0 & 0 &  \cr
\quad \vdots  & &  &  &  & \vdots & \cr
\quad 0 & 0 & 0 & \cdots  & 0 & 0 &\cr
\end{matrix}
\end{pmatrix},
\]
where $\ast$ are arbitrary elements in $\cc$ and $\diamond$ are arbitrary
elements in $\rr^+$. Lets call this slice $\ms$.

\section{The exact slice in the other cases (lining up with $\ms\mun(m,n)$)}

In the analysis of other two non-trivial cases in which $M$ is nontrivial we
may exploit the results obtained in the previous section by embedding
appropriately $\ms\mo(m,n)$ and $\ms\mp(2m,2n)$ in, respectively,
$\ms\mun(m,n)$ and $\ms\mun(2m,2n)$ (lining up with $\ms\mun(m,n)$).

\m In the case $G_0 = SO(m,n)^0$ we have $\ms\mo(m,n) = \mk_2 \oplus \mp_2$,
where
\[
\mk_2 = \ms\mo(m,n) \cap \mk_1 =  \ms(\mo(m) \times \mo(n))
\]
and
\[
\mp_2 = \ms\mo(m,n) \cap \mp_1 = \left \{ \begin{pmatrix} 0 & B \cr B^T & 0
\end{pmatrix} \vert B \in M_{m \times n}(\rr) \right\}.
\]
As a maximal abelian subspace of $\mp_2$ we can choose the same as for
$\mp_1$. In particular, we have
\[
M_2 = Z_{K_2}(\ma) = M_1 \cap K_2 = \left \{ \begin{pmatrix} A & 0 \cr 0 & 0
\end{pmatrix} \vert A \in SO(m-n) \right \} \simeq SO(m-n).
\]
Thus in this case the slice is given by matrices of the form
\[
\begin{pmatrix}
\begin{matrix}
0 & \cdots & 0 &  \ \cr
\vdots &  \ddots &  \vdots  & \ \cr
0 & \cdots & 0 &  \ \cr
\cr
\cr
\end{matrix}
&
\begin{matrix}
\quad 0 & 0 & \hdots & \hdots & 0 & 0 & \quad  \vline \cr
\quad \vdots  & &  &  &  & \vdots & \quad \vline \cr
\quad 0 & 0 & \hdots & \hdots  & 0 & 0 & \quad \vline \cr
& & & & & & \quad \vline \cr
 \cr
\end{matrix}
&
\begin{matrix}
\ast & \cdots &  \cdots &  \cdots &\ast \cr
& \ddots &  &   & \vdots \cr
0 &  &\ast & \cdots & \ast\cr
\cr
\hline \cr
\end{matrix}
\cr
\begin{matrix}
0 & \cdots & 0 &  \ \cr
\vdots &  \ddots &  \vdots  & \ \cr
0 & \cdots & 0 &  \ \cr
\vdots &  \ddots &  \vdots  & \ \cr
0 & \cdots & 0 &  \ \cr
\cr
\hline
\end{matrix}
&
\begin{matrix}
\quad 0 & 0 & \hdots & \hdots & 0 & 0 & \quad \vline \cr
\quad \vdots  & &  &  &  & \vdots &\quad \vline \cr
\quad 0 & 0 & \hdots &\hdots & 0 & 0 & \quad \vline \cr
\quad \vdots  & &  &  &  & \vdots & \quad \vline \cr
\quad 0 & 0 & \hdots & \hdots  & 0 & 0 & \quad \vline \cr
& & & & & & \quad \vline \cr
\hline
\end{matrix}
&
\begin{matrix}
\ast & \cdots & \cdots & \ast & 0 \cr
\vdots  & \ddots  & \ast & \cdot & \ast \cr
\vdots & \ast & \cdot & \ast & \vdots \cr
\ast & \cdot & \ast &  & \vdots \cr
0 & \ast & \cdots & \cdots  & \ast  \cr
\cr
\hline
\end{matrix}
\cr
\begin{matrix}
\ast &  & 0 & \vline \cr
  \vdots & \ddots &  & \vline \cr
 \vdots &  & \ast & \vline\cr
 \vdots & & \vdots & \vline\cr
 \ast & \cdots & \ast & \vline
\end{matrix}
&
\begin{matrix}
\ast & \cdots & \cdots & \ast & 0 \cr
\vdots  & \ddots  & \ast & \cdot & \ast \cr
\vdots & \ast & \cdot & \ast & \vdots \cr
\ast & \cdot & \ast &  & \vdots \cr
0 & \ast & \cdots & \cdots  & \ast  \cr
\end{matrix}
&
\begin{matrix}
\cr
\quad 0 & \cdots & \cdots  & \cdots & 0  \cr
 \quad \vdots  &   &  &  & \vdots  \cr
 \quad 0 &  \cdots  & 0 & \cdots  &  0 \cr
\quad \vdots  &   &  && \vdots \cr
\quad 0 &  \cdots  & 0 & \cdots & 0 \cr
\end{matrix}
\end{pmatrix},
\]
where $\ast$ are arbitrary elements in $\rr$.

\m
In an analogous way we compute the slice for the group $Sp(2m,2n)$. Obviously
we have to do the lining up with $\ms \mun(2m,2n)$ which we also call $\mg_1$.
We also use the same notation for the Cartan decomposition and the maximal
Abelian subspaces as for $\ms\mun(m,n)$. The
Cartan decomposition of $\mg_3 = \ms\mp(2m,2n)$ is given by  $\mg_3 = \mk_3
\oplus \mp_3$, where
\[
\mk_3 = \mg_3 \cap \mk_1 = \mun\ms\mp_m \oplus \mun\ms\mp_n
\]
and
\[
\mp_3 = \mg_3 \cap \mp_1 = \left \{
\begin{pmatrix} 0 & B \cr \bar B^T & 0 \end{pmatrix} \in \mg_3 \mbox{ with }
B \in M_{2m \times 2n}(\cc) \right \}.
\]
Unfortunately the maximal abelian subspace $\ma_1$ of $\mp_1$ does not contain
a maximal abelian subspace of $\mp_3$. Therefore we choose the following
maximal abelian subspace $\ma_1^\prime$ of $\mp_1$,
\[
\ma_1^\prime := \left \{  X=
\begin{pmatrix}
0 & B \cr
\bar B^T & 0
\end{pmatrix}
\mbox{ with } B=
\begin{pmatrix}
\vdots & \cdots & \vdots \cr
0 & \cdots & 0 \cr
0 &  \cdots & a_{2n} \cr
0 & \cdot & 0  \cr
a_1 & \cdots & 0  \cr
0 & \cdots  & 0 \cr
\vdots & \cdots & \vdots
\end{pmatrix} \in M_{2m \times 2n}(\cc),
\ a_j \in \rr \right\},
\]
where $a_{2n}$ is contained in the $(m-n+1)$th row of the matrix $B$. Then
$M_1^\prime = Z_{K_1}(\ma_1^\prime)$ consits of all matrices of the form
\[
\begin{pmatrix}
U_1 & 0 & U_2 & 0 \cr
0 & D_1 & 0 & 0 \cr
U_3 & 0 & U_4 & 0 \cr
0 & 0 & 0 & D_2
\end{pmatrix} \in S(U(m) \times U(n)) \mbox{ with } \begin{pmatrix} U_1 & U_2 \cr U_3 & U_4
\end{pmatrix} \in U(2m-2n),
\]
$  D_1 = diag(e^{it_1},...,e^{it_{2n}})$
and $D_2 = diag(e^{it_{2n}},...,e^{it_1})$. Now one can choose a maximal
abelian subspace $\ma_3$ in $\mp_3$ which is contained in $\ma_1^\prime$,
namely
\[
\ma_3 = \left \{ X = \begin{pmatrix} 0 & B \cr B^T & 0 \end{pmatrix}
\in \ma_1^\prime \mbox{ with }  a_i  =  a_{i+n} \in \rr, 1\leq i \leq n
\right \},
\]
and then
$M_3
= M_1^\prime \cap K_3$ consists of all matrices of the form
\[
 \begin{pmatrix}
U_1 & 0 & 0 & 0 \cr
0 & D_1 & 0 & 0 \cr
0 & 0 & \bar U_1 & 0 \cr
0 & 0 & 0 & D_2
\end{pmatrix} \in USp(m) \times USp(n) \mbox{ with } U_1 \in U(m-n),
\]
$  D_1 = diag(e^{it_1},...,e^{it_{n}},e^{it_1},...,e^{it_n})$
and $D_2 = diag(e^{it_{n}},...,e^{it_1},e^{it_n},...,e^{it_1})$ .

\section{Poisson Structure}

We use the exact slice to compute the Poisson structure with respect to the
new coordinates which are given as follows. Let $Q = (q_1,...,q_N)$ denote the
coordinates for $\ma$ in the first factor of the product
$S = \ma \times \ma \oplus \ms$ regarded as an $K$-invariant function on
$\mp \times \mp$. Further let $V$ denote the standard matrix coordinates of
the second factor of $S$ again regarded as an invariant function on
$\mp \times \mp$. Finally for $x = k \cdot s$ with $s \in S$ and
$x  \in \mp \times \mp$ we define $U(x) \in K$
to be the matrix given by $k$. Define $\mathrm d U$ to be the matrix of
$1$-forms $\mathrm d U_{ij}$ and $W := U\inv \mathrm d U$.
Note that $W = - W^\dagger$ since $U^\dagger U = Id$. This also gives
$\mathrm d W = - W \land W$.

\m
We now compute the symplectic form $\omega = \re (\tr (\mathrm d X \land
\mathrm dY)$ using these invariant functions. For this we write $\omega =
\mathrm d \theta$, where $\theta = \re (\tr(Y \mathrm dX))$. We have thus
\[
\begin{aligned}
\theta & =  \re(\tr( U V U\inv \mathrm d(U Q U\inv)))  \cr
& = \re (\tr ( UVU\inv \mathrm d U Q U\inv
+ \re(\tr( UVU\inv U \mathrm d Q U\inv))
+ \re(\tr( UVU\inv U Q \mathrm d U\inv)) \cr
& = \re(\tr(VW Q))
+ \re(\tr(V \mathrm dQ))
+ \re(\tr(VQ W\inv) \cr
& = - \re(\tr(VW\inv Q))
+ \re(\tr(V \mathrm dQ))
+ \re(\tr(VQ W\inv) \cr
& = \re(\tr(V\mathrm d Q))  + \re(\tr(V [Q,W\inv]))).
\end{aligned}
\]
We can now simplify the second summand,
\[
\begin{aligned}
\re(\tr(V [Q,W\inv])))
& = - \re(\tr([V,Q]W)) \cr
& = - \frac 12 \tr ([V,Q]W + \overline{[ V,  Q]}\bar W) \cr
& = - \frac 12 \tr( [V,Q]W + [V,Q]^T W^T) \cr
& = - \tr ([V,Q]W) \cr
& = - \tr(l W), \cr
\end{aligned}
\]
where we again use $\mk_0 \subset \ms\mun(N)$.
Therefore we have
\[
\omega = \mathrm d \theta = \re(\tr(\mathrm dV \land \mathrm dQ))
- \tr (\mathrm d l \land W) + \tr (l W \land W).
\]
Due to the structure of $\ms$ we can replace $\mathrm dV$ by its $\ma$-part
with respect to the decomposition $\ma \oplus \ms$ which we call $\mathrm d P$.
Moreover, since $Q$ and $P$ are real
we have
\[
\omega = \tr(\mathrm d P \land \mathrm d Q)
- \tr (\mathrm d l \land W) + \tr (l W \land W),
\]
which shows that the Poisson structure splits with the pair $(P,Q)$ being
canonical, commuting with $l$. We want show now that $l$  has the
coadjoint Poisson structure of $\mk_0$.

\begin{proposition}
The map $l : \mp \times \mp \to \mk^\ast$ is a Poisson morphism.
\end{proposition}

We regard $l$ as a complex matrix valued map with values in $\ms(\mun(m)
\times \mun(n))$ and let $\mathrm d l = (\mathrm d l_{ij})$ be the matrix
of $\cc$-valued $1$-forms.

\m
We now compute the Hamilton vector field $V_f$ of a function $f = f(l)$
of $l$ alone. It is defined by the equation $\mathrm d f(Z) = \omega(V_f,Z)$
for any real valued field $Z$ on $\mp \times \mp$. Since the symplectic forms
splits an $f$ is a function of $l$ alone
we only have to consider the pieces of the field $Z$ which involve
$(\partial/\partial l)$ and $(\partial/ \partial W)$. Therefore we have
\[
\omega(V_f,Z) = - \tr(\mathrm d l \land W - l W \land W)(A,B),
\]
where $A = \tr(V_f^l (\partial/ \partial l)^T)
+ \tr(V_f^W(\partial/\partial W)^T)$ and
$B = \tr(Z^l (\partial/\partial l)^T) + \tr(Z^W(\partial/ \partial W)^T)$.
Then we have
\[
\begin{aligned}
\omega(V_f,Z) &= - \tr(V_f^l Z^W + Z^l V_f^W
+ l V_f^W Z^W - l Z^W V_f^W) \cr
& =  \tr(V_f^l Z^W + Z^l V_f^W + [l,V_f^W]Z^W).
\end{aligned}
\]
This implies
\[
\mathrm d f(Z) = \tr(Z^l(\partial f/\partial l)^T = \tr(Z^l V_f^W)
+ \tr(Z^W (-V_f^l + [l,V_f^W]),
\]
and, therefore, $V_f^W = (\partial f/\partial l)^T$ and
$V_f^l = [l,(\partial f/ \partial l)^T]$.

\section{The cotangent bundle}

The moment map on the cotangent bundle $N:= T^\ast Q$ is given by
\[
\mu \colon T^\ast Q \to \mk^\ast, \quad \alpha \mapsto (\xi \to
\alpha(\xi_X(\pi(\alpha)).
\]
After identifying $T^\ast Q$ with $\mp \times \mp$  we have
\[
\mu(X_1,X_2)(\xi) = B(X_2,[\xi,X_1]),
\]
where $B$ is as above. Using the $K$-invariance of $B$ we get
\[
\mu(X_1,X_2)(\xi) = B(X_2,[\xi,X_1]) = -B(X_2,[X_1,\xi]) = B([X_1,X_2],\xi).
\]
With the identification $\mk^\ast \simeq \mk$ by the inner product $-
B\vert_{\mk \times \mk}$ this gives
\[
\mu(X_1,X_2) = [X_2,X_1].
\]
Define $l \colon S \to \mz_\mk(\ma)^{\sperp_\mk}$ by $ (q,p,r) \mapsto
\mu(q,p,r) = [r,q]$. Since $\mp = K \acts \ma$ and $Z_K(\ma)$ fixes $\ma$
pointwise the map
\[
f \colon K/Z_K(\ma) \times \ma \to \mp, ([k],\xi) \mapsto \Ad(k) \xi,
\]
is well defined and surjective. Let $E$ be a generic point of $\ma$, e.\,g.\
$E$ is the half sum of positive roots. Since the stabilizer of $K$ in $E$ is
precisely $Z_K(\ma)$ the derivative of $f$ in the point $([e],E)$, which is
given by
\[
\begin{aligned}
Df([e],E) : T_{([e],E)}(K/Z_K(\ma) \times \ma) \simeq \mk/\mz_\mk(\ma) \times
\ma \to & \ \mp, \cr  ([\xi], \eta) \ \  \mapsto & \ \xi_X(E) + \eta = [\xi,E]
+ \eta,
\end{aligned}
\]
is an isomorphism of $Z_K(\ma)$-representations. Using the inner product
$-B\vert_{\mk \times \mk}$ we can identify $\mk/\mz_\mk(\ma)$ with
$\mz_\mk(\ma)^\sperp$ and we get an isomorphism of the
$Z_K(\ma)$-representation spaces $\mz_\mk(\ma)^\sperp $ and $\ma^\sperp$ given
by $\xi \mapsto [\xi,E]$.

\begin{remark}
Let $q \in \ma$. Then the linear map $ A_q:=\ad(E) \circ \ad(q) \colon
\mz_\mk(\ma)^\sperp \to \mz_\mk(\ma)^\sperp$ is selfadjoint since
\[
\kappa(A_q(\xi),\eta) = \kappa([E,[q,\xi],\eta]) = \kappa(\xi,[q,[E,\eta]]) = \kappa(\xi,[E,[q,\xi]])
= \kappa(\xi,A_q(\eta))
\]
for $\xi,\eta \in \mz_\mk(\ma)^\sperp$. In particular, $A_q$ is diagonalizable.
\end{remark}

\n
Let $e_j$ be a basis for $\mz_\mk(\ma)^\sperp $ consisting of eigenvectors
of $A_q$ with eigenvalues $\lambda_j(q)$.

\m
The same is true if we regard $A_q$ as a map from $\ma^\sperp$ to itself.





\section{Slice densities}

\m Let $\mathrm d \lambda_M$ denote the Liouville measure associated with the
Liouville form $\omega_N := \omega^n$ on $N$. Further let $\omega_S$ denote the
linear  volume form on the slice $S$. Then the slice density $\rho : S \to
\rr^{\geq 0}$ is given by the equation
\[
\int_N f \ \omega_N = \int_S (f \cdot \rho) \ \omega_S
\]
for all compactly supported functions $f \in \mathcal E_0(M)^K$. Our goal is to
compute this slice density and to prove the following.

\begin{proposition}\label{slice denity(p)}
The canonical slice measure is given by $\rho \ \mathrm d \lambda_S = \mathrm d q \ \mathrm d p \ \mathrm d l$
\end{proposition}

\m Let $\alpha \colon K \times S \to M, (k,x) \mapsto k \acts x$ denote the
action map. This is a $Z_K(\ma)$-principal bundle and we can compute the slice
density as follows.

\m Let $\omega_K$ the invariant volume form on $K$ normalized by $\int_K
\omega_K = 1$ and let $\omega_S$ the standard Euclidean volume form $\mathrm d
q \land \mathrm d p \land \mathrm d r$ on $S$. Let $\mathcal T$ be the
invariant frame field along the fibers of $\alpha$. Then we define the function
$\rho \colon K \times S \to \rr^{\geq0}$ by the equation
\[
\rho \ \iota_{\mathcal T}(\omega_K \land \omega_S) = \alpha^\ast \omega_M
\]
where $\iota_{\mathcal T}$ denotes contraction with the frame $\mathcal T$.
Since all of the differential forms which are involved are invariant under the
group $K$, the function $\rho$ is also $K$-invariant and therefore defines a
function on the slice $S$ which we also denote by $\rho$.

\m
Applying Fubini's Theorem we get
\[
\int_{K\times S} \alpha^\ast(f) \ \rho \  \omega_K \land \omega_S
= \int_S \Bigl( \int_K \rho \alpha^\ast (f) \ \omega_K \Bigr ) \omega_S
=\int_S f \rho \ \omega_S
\]
for any $f \in \mathcal E_0(M)^K$. Let $\omega_{Z_K(\ma)}$ denote the invariant
volume form on $Z_K(\ma)$ normalized by $\int_{Z_K(\ma)}\omega_{Z_K(\ma)} = 1$
such that
\[
\omega_{Z_K(\ma)} \land \iota_{\mathcal T}(\omega_K \land \omega_S) = \omega_K \land \omega_S.
\]
Then by fiber integration
we get
\[
\begin{aligned}
\int_{K\times S}\alpha^\ast(f) \rho \ \omega_K \land \omega_S = & \int_{K\times
S} \alpha^\ast(f) \ \rho \ \omega_{Z_K(\ma)} \land \iota_{\mathcal T}(\omega_K
\land \omega_S) \cr & \cr = & \int_{K \times S} \omega_{Z_K(\ma)} \land
\alpha^\ast( f \, \omega_M) \cr & \cr = & \int_M \biggl( \int_{Z_K(\ma)}
\omega_{Z_K(\ma)} \biggr) f \, \omega_M = \int_M f \, \omega_M. \cr
\end{aligned}
\]
This shows that the function $\rho$ is the slice density defined above. In the
following we want to compute $\rho$ in a more explicit way. At $s =(q,p,l)\in
S$, we compute the determinant of the projection of the map
\[
\alpha_\ast : \mz_\mk(\ma)^\sperp \to \mp \times \mp \simeq \mz_\mk(\ma)^\sperp
\times S, \quad \xi \mapsto \alpha_\ast(\xi),
\]
onto the factor $\mz_\mk(\ma)^\sperp$. It can be computed as follows.
\[
\alpha_\ast(\xi) = \ddt \alpha(\exp(t \xi),s)
 =  \ddt \Ad(\exp(t\xi))(s)
 = [\xi,s] = ([\xi,q],[\xi,(p,l)]),
\]
with $[\xi,q] \in \ma^{\sperp} \simeq \mz_\mk(\ma)^{\sperp}$.
Then we have
\[
Pr(\alpha_\ast(\xi))= Pr([\xi,s])  = [\xi,q].
\]
In particular, we get
\[
Pr(\alpha_\ast(e_j)) = \lambda_j(q) e_j,
\]
which gives
\[
\rho(s) = \Bigl\vert \prod_{j} \lambda_j(q) \Bigr\vert.
\]

\m {\it Proof of Proposition \ref{slice denity(p)}}.
\ Let $(q,p,r)$ denote
the linear coordinates on $S = \ma \times \mp$. Then $\omega_S = \mathrm d q
\land \mathrm d p \land \mathrm d r$ is the linear volume form on $S$. Consider
the coordinate change
\[
\varphi : S \to S , \quad (q,p,r) \mapsto (q,p,l):= (q,p,\mu(q,r)) =
(q,p,[r,q]).
\]
By the transformation rule we get
\[
\mathrm dq \land \mathrm dp \land \mathrm dl = \vert \mathrm D\varphi(q,p,r)
\vert \cdot \mathrm dq \land \mathrm dp \land \mathrm dr = \Bigl \vert \prod_j
\lambda_j(q) \Bigr \vert \cdot  \omega_S = \rho \ \omega_S.
\]
\qed

\m As an example let us compute explicitly the slice density for the symmetric space
$SU(m,n)/S(U(m) \times U(n))$

\m We consider the following basis of $\mz_\mk(\ma)^\sperp$. Let $e^k_{i,j},k
=1,2; 1 \leq i \leq m-n; 1 \leq j \leq n$ and $i<j$ denote the matrices of the
form
\[
\begin{pmatrix}
0 & B & 0 \cr
-\bar B^T & 0 & 0 \cr
0 & 0 & 0
\end{pmatrix} \mbox{ with } B =(b_{k,l}) \in M_{(m-n)\times n}(\cc) \hskip 4cm
\]
\[
\hskip 4cm \mbox{ and } b_{k,l} = \begin{cases}
1 \qquad  \mbox{for }  (k,l)=(i,j) \mbox{ and } k = 1 \cr
i \qquad  \mbox{for }  (k,l)=(i,j) \mbox{ and } k = 2 \cr
0 \qquad \mbox{otherwise.} \end{cases}
\]
Then $A_q \cdot e^k_{i,j} = q_j \cdot e^k_{i,j}$. Further let $e_i, 1 \leq i
\leq n$ denote the matrices of the form
\[
\begin{pmatrix}
0&0&0 \cr
0&C&0 \cr
0&0& C^\prime
\end{pmatrix}, \mbox{ where } C= diag(c_1,...,c_n) \mbox{ with }
c_l = \begin{cases} i \qquad  \mbox{ for } l=i \cr 0 \qquad \mbox{otherwise,} \end{cases}
\]
and $C^\prime = diag(-c_q,...,-c_1)$. Then $A_q \cdot e_i = q_i \cdot e_i$.
We further define the basis elements $f^{\pm}_{i,j}, 1 \leq i < j \leq n$
to be the matrices
\[
\begin{pmatrix}
0 & 0 & 0 \cr
0 & D & 0 \cr
0 & 0 & D^\prime
\end{pmatrix} \mbox{ where } D = (d_{kl}) \in M_{n \times n}(\cc) \mbox{ with }
d_{kl} = \begin{cases} 1 \qquad \mbox{ for } (k,l)=(i,j) \cr
-1 \quad \mbox{ for } (k,l) = (j,i) \cr
0 \qquad \mbox{ otherwise,} \end{cases}
\]
and $D^\prime$ is equal to $\pm D$ reflected at the anti-diagonal. For these
basis elements we have $A_q \cdot f^\pm_{i,j} = (q_i \pm q_j) f^\pm_{i,j}$. The
last basis elements $g^\pm_{i,j}, 1 \leq i < j \leq n$ are given by matrices of the form
\[
\begin{pmatrix}
0 & 0 & 0 \cr
0 & D & 0 \cr
0 & 0 & D^\prime
\end{pmatrix} \mbox{ where } D = (d_{kl}) \in M_{n \times n}(\cc) \mbox{ with }h
d_{kl} = \begin{cases} i \qquad \mbox{ for } (k,l)=(i,j) \cr
i \qquad \mbox{ for } (k,l) = (j,i) \cr
0 \qquad \mbox{ otherwise,} \end{cases}
\]
and $D^\prime$ is again $\pm D$ reflected at the anti-diagonal.
We have $A_q \cdot g^\pm_{i,j} = (q_i \pm q_j) \cdot g^\pm_{i,j}$.

\m
Therefore the slice density is given by
\[
\rho(s) = \prod_{i=1}^n q_i^{2(m-n)+1} \cdot \prod_{i<j}(q_i^2-q_j^2)^2.
\]

\m
By an analogous computation for $SO(m,n)^0/S(O(m) \times O(n))$ we get the
slice density
\[
\rho(s) =  \Bigl \vert \prod_{i=1}^n q_i^{m-n} \cdot \prod_{i<j}(q_i^2-q_j^2) \Bigr \vert
\]

\m
{\bf Remark.} Note that the density $\rho (s)$ only depends on
the variable $q$, i.e., on the eigenvalues of the operators
at hand. When formulated in our notation, the usual procedure 
in random matrix theory is to start with a $K$--invariant
probability density $d$ on the cotangent bundle of the symmetric
space so that the resulting density on the slice defines a 
probability measure $d\rho dqdpdr$. In classical examples
where $M$ is not present it is usually a simple matter to compute the image
``spectral'' measure on $\ma$ or on the Weyl chamber $\ma_+$. It 
would be interesting to know if the presence of $M$ is of physical
interest, e.g., if it would be appropriate to simply take
the standard norm function and use $K$-invariant probability 
distribution $\exp(-\frac12 \Vert\cdot \Vert^2) \ \mathrm d \lambda_M$. 

\section {Other symmetric spaces}

{\bf Symmetric spaces of Type II.}

\bigskip\noindent
Above we restricted our discussion to symmetric spaces of
simple Lie groups which are \emph{not} complex. If the real
group $G$ happens to be complex, one refers to the
associated symmetric space $G/K$ as being of Type II. A
typical example is $\mathrm{SL}_n(\mathbb C)/\mathrm{SU}_n$.
Actually the above discussion simplifies in this situation.
The point is that if $G$ is complex, then the subgroup $K$ is 
a compact real form and at the Lie algebra level
$\mathfrak p=i\mathfrak k$. If $\mathfrak t$ is the
Lie algebra of a maximal torus in $\mathfrak k$, then
$\mathfrak a:=i\mathfrak t$ is a maximal Abelian subspace
of $\mathfrak p$.  Since the centralizer of $\mathfrak t$
in $\mathfrak k$ is $\mathfrak t$ itself, it follows that
$\mathfrak m=0$.

For example, in the case of $G=\mathrm{SL}_n(\mathbb C)$ the Cartan decomposition
of a matrix in $\mathfrak g=\mathfrak k\oplus \mathfrak p$ amounts to writing
it as the sum of antihermitian and hermitian matrices. Hence,
our work above just puts the classical level dynamics discussion 
(\cite {haake00},\cite {hzkh01}) for pairs of hermitian matrices 
in a symmetric space framework.  

\bigskip\noindent
{\bf Symmetric spaces of compact type}

\bigskip\noindent
If $G$ is compact and $G/K$ is a symmetric space, e.g., the
Grassmannian $\mathrm{Gr}_q(\mathbb C^n)$ of q--dimensional
complex subspaces in $\mathbb C^n$, then one can also discuss
level dynamics in a setup similar to that above.  Conceptually
it is convenient to think about this in a situation where
the the duality between symmetric spaces of compact and noncompact
type is visible.  For this it is convenient to introduce 
a bit of notation.  Details of the below discussion can
be found in (\cite{FHW}).

\bigskip\noindent
If $G_0$ is a simple Lie group of noncompact type with a
given Cartan decomposition $\mathfrak g_0=\mathfrak k_0+\mathfrak p_0$,
we consider the complexification 
$\mathfrak g:=\mathfrak g_0+i\mathfrak g_0$ and the associated
complex semisimple Lie group $G$.  Note that in the Type II
case mentioned above $\mathfrak g$ is the direct sum of
two copies of $\mathfrak g_0$.  Otherwise, $\mathfrak g$ is
also simple.

One observes that $\mathfrak g_u=\mathfrak k_0+i\mathfrak p_0$
is a compact real form of $\mathfrak g$. So, going to the
Lie group level, we have the complex group $G$ containing
the noncompact real form $G_0$ and the compact real form $G_u$.
Now let $K_0$ be the maximal compact subgroup of $G_0$ which
is associated to $\mathfrak k_0$ and $K$ the complex
subgroup of $G$ which is associated to 
$\mathfrak k=\mathfrak k_0+i\mathfrak k_0$.  If $x_0$
is the neutral point in the complex (affine) symmetric space
$G/K$, then $G_0.x_0=G_0/K_0$ is initial symmetric space
of noncompact type and $G_u.x_0=G_u/K_0$ is the dual symmetric
space of compact type.  The cotangent space at the neutral
point of the noncompact symmetric space is $\mathfrak p_0^*$
and that of the compact symmetric space is $i\mathfrak p_0^*$.

Above in the case of noncompact symmetric spaces we have
use the fact that using the exponential map we may
identify the given symmetric space with $\mathfrak p_0$.
In particular the cotangent bundle is trivial.  This
is essentially never the case for compact symmetric spaces,
e.g., almost no spheres have this property.  Furthermore,
the exponential map $\mathrm{exp}:i\mathfrak p_0\to G$
is not as simple in this case. The difficulty can, however, be
isolated in the maximal Abelian subspace $i\mathfrak a_0$
whose associated group is a compact torus. Here
$\mathrm{exp}:i\mathfrak a_0\to A$ is nothing other than
the usual covering mapping which amounts to dividing out
a vector space by a lattice of periods.

Using this and the fact that (modulo a certain Weyl group)
$i\mathfrak a_0$ is a slice for the $K_0$--action on
$\mathfrak p_0$, one observes that, after going to 
the complement $i(\mathfrak p_0)_{\mathrm{gen}}$ of
an approriate set of measure zero in $\mathfrak p_0$,
we have an identification of the phase space
at hand with the product 
$i(\mathfrak p_0)_{\mathrm{gen}}\times i\mathfrak p_0^*$.
All of the above considerations for symmetric spaces 
of noncompact type can now be carried out on this
set of \emph{generic points} in the cotangent bundle
of the compact symmetric space.

The above indicates that obtaining coordinates
for considerations of level dynamics in the
cotangent bundle of symmetric space of compact
type is a more difficult matter than in the 
case of noncompact symmetric spaces.  On the
other hand, the compact
symmetric space has one major advantage:  the
complex symmetric space $G/K$ is naturally 
identifiable with its cotangent bundle. In other
words, the relevant phase space is itself
a complex symmetric space.  This is not
the case for the noncompact symmetric space.
There is indeed a map from its cotangent
bundle into $G/K$ (polar coordinates), but
this degenerates at a certain point. There
is, however, a precisely defined maximal neighborhood of
the zero--section of this phase where the
polar coordinate mapping is a diffeomorphism
onto its image $\mathcal U$ in $G/K$.  Thus
on $\mathcal U$, where perfect coordinates
and natural invariant measures are available, 
it is possible to consider the level dynamics
related to both the compact and noncompact 
symmetric spaces.   .


\section{Acknowledgments}
The support by SFB/TR12 `Symmetries and Universality in Mesoscopic Systems'
program of the Deutsche Forschungsgemeischaft and Polish MNiSW grant No
1P03B04226 is gratefully acknowledged.

\end{document}